\documentclass[global,twocolumn]{svjour}
\usepackage{graphicx}
\usepackage{siunitx}
\usepackage{braket}
\usepackage{lineno}
\usepackage{ulem} 
\usepackage{url}
\begin{document}
\title{Autonomous frequency stabilization of two extended cavity diode lasers at the potassium wavelength on a sounding rocket}
\author{Aline N. Dinkelaker\inst{1} \and Max Schiemangk\inst{1} \and Vladimir Schkolnik\inst{1} \and Andrew Kenyon\inst{1} \and Kai Lampmann\inst{2} \and Andr\'{e} Wenzlawski \inst{2}  \and Patrick Windpassinger\inst{2} \and Ortwin Hellmig\inst{3} \and Thijs Wendrich\inst{4}  \and Ernst M. Rasel\inst{4} \and Michele Giunta\inst{5,6} \and Christian Deutsch\inst{5} \and Christian K\"urbis\inst{7} \and Robert Smol\inst{7} \and Andreas Wicht\inst{1,7} 
\and Markus Krutzik\inst{1} \and Achim Peters\inst{1,7}
}                     
%
%
\institute{Institut f\"ur Physik, Humboldt-Universit\"at zu Berlin, 12489 Berlin, Germany\\ \email{aline.dinkelaker@physik.hu-berlin.de} 
\and Institut f\"{u}r Physik, Johannes Gutenberg-Universit\"{a}t Mainz, 55099 Mainz, Germany 
\and Institut f\"{u}r Laserphysik, Universit\"{a}t Hamburg, 22761 Hamburg, Germany 
\and Institut f\"ur Quantenoptik, Leibniz Universit\"at Hannover, 30167 Hannover, Germany
\and Menlo Systems GmbH, 82152 Martinsried, Germany
\and Max Planck Institut f\"ur Quantenoptik, 85748 Garching, Germany
\and Ferdinand-Braun-Institut, Leibniz-Institut f\"ur H\"ochstfrequenztechnik, 12489 Berlin, Germany
}
\date{}
%
\maketitle
\begin{abstract}
We have developed, assembled, and flight-proven a stable, compact, and autonomous extended cavity diode laser (ECDL) system designed for atomic physics experiments in space. To that end, two micro-integrated ECDLs at 766.7~nm were frequency stabilized during a sounding rocket flight by means of frequency modulation spectroscopy (FMS) of $^{39}$K and offset locking techniques based on the beat note of the two ECDLs. The frequency stabilization as well as additional hard- and software to test hot redundancy mechanisms were implemented as part of a state-machine, which controlled the experiment completely autonomously throughout the entire flight mission. 
\end{abstract}

\section{Introduction}
Laser systems are a key technology for space-borne applications in optical communication, time keeping, inertial navigation, and geodesy \cite{Schlepp:2014,Cacciapuoti:2009,Prestage:2007,Yu:2006,Chou:2010,Carraz:2014}, as well as high precision measurements in fundamental physics, such as clock comparisons~\cite{Cacciapuoti:2009,Leveque:2014,Leveque:2015}, gravitational wave detection~\cite{Hogan:2011,Graham:2013}, and cold atom based quantum sensors~\cite{Aguilera:2014,Schkolnik:2016}. The laser system described in this paper is motivated by the developments and demands of cold atom based quantum sensors for tests of the Einstein equivalence principle (EEP)~\cite{Schlippert:2014,Zhou:2015,Bouyer:2008} on future satellite missions (e.g.~\cite{Aguilera:2014,Schuldt:2015}). In this context, laser light is required for optical cooling, trapping, manipulation, and detection of the atoms. This poses high demands on the laser system's functionality, specifically on laser frequency, intensity, their corresponding stability, and tuning range. Additional requirements on mechanical stability, weight, and size of the hardware arise from the deployment to space itself. Beyond that, redundancy and autonomy features should be implemented due to restricted access to the experiment.  

Before reaching out to space, any experiment and its subsystems have to be qualified in terms of technological readiness by means of environmental testing. Additionally, experimental sequences have to be developed and parameters optimized in a scenario that resembles the target environment. In the context of quantum sensors for EEP tests in space, this includes microgravity. On ground, these quantum sensors can be tested in microgravity at a drop tower, e.g. the ZARM Bremen drop tower~\cite{Zoest:2010,Herrmann:2012,Muntinga:2013}, where microgravity ($< 10^{-6}$~\textit{g}) of several seconds (up to 9.6~s) can be achieved. Airborne tests can be conducted on parabolic flights~\cite{Bouyer:2008}, where longer times (about 20~s per parabola on-board a Novespace A300 Zero-\textit{g} airplane) of reduced gravity ($< 0.05$~\textit{g}) are possible. For an environment that is closer to conditions on a satellite or the International Space Station (ISS), sounding rockets are the ideal test bed, providing several minutes in reduced gravity better than $10^{-4}$~\textit{g}. They are also a suitable preparation for a future transport of the experiment to Earth-orbit platforms. 

\begin{figure}
	\centering
		\includegraphics[width = 0.95\columnwidth]{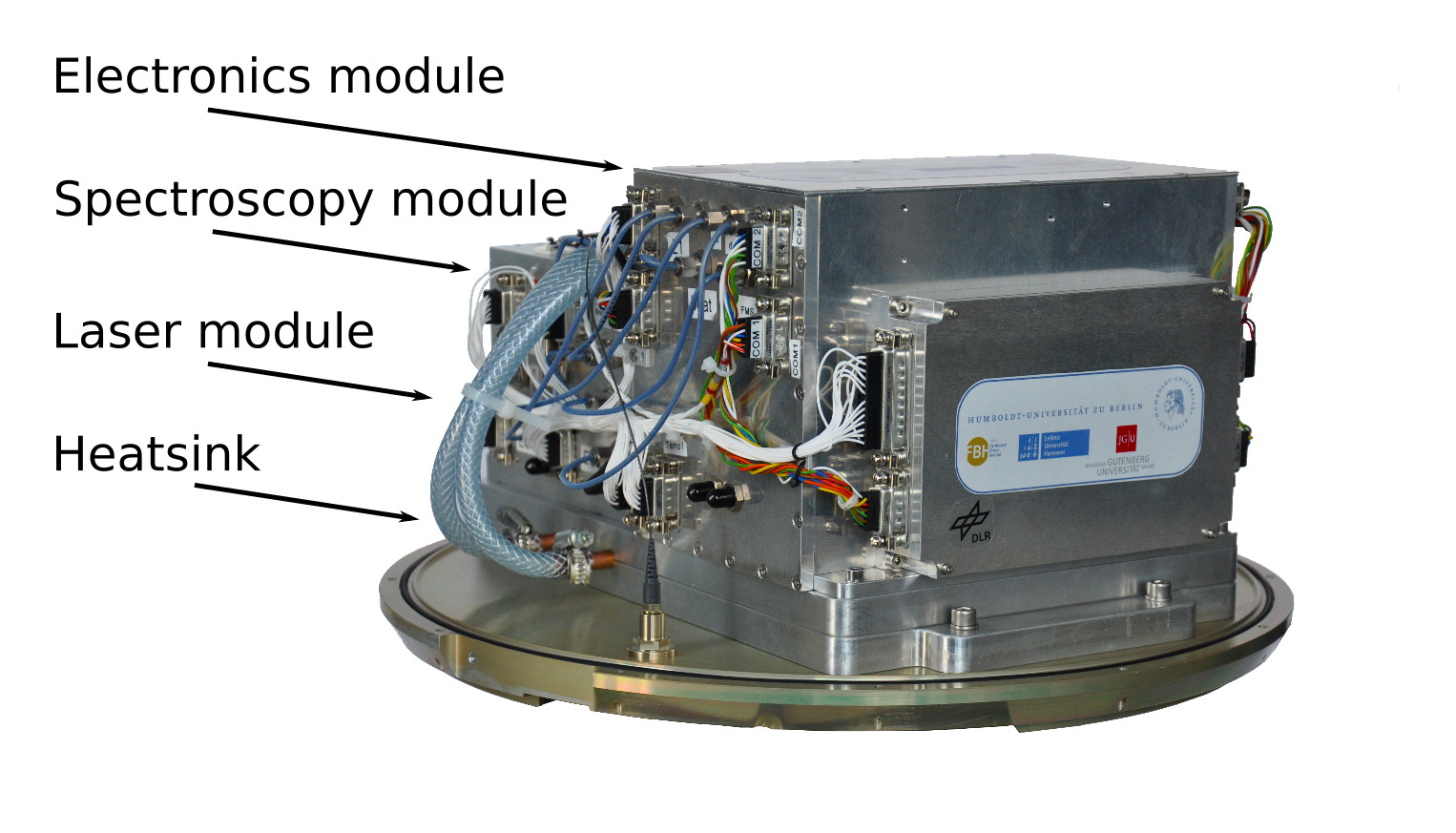}
	\caption{The KALEXUS payload has a size of $345 \times 218 \times 186$~mm$^3$ (L~$\times$~W~$\times$~H) and a mass of 16~kg. It consists of three different modules integrated into individual, space-grade aluminum housings, connected via optical and electrical interfaces, and mounted onto a heatsink. During flight, the experiment is heat shielded and pressurized within a dome (not shown).}
	\label{fig:KALEXUS_beschriftet}
\end{figure}

The KALEXUS (German acronym for \textit{Kalium Laser Experimente Unter Schwerelosigkeit}) experiment tests a frequency stabilized semiconductor diode laser system on-board the TEXUS~53 sounding rocket. Semiconductor diode lasers are chosen as light sources as they are compact, robust, and available at a variety of wavelengths. This enables complex systems with a high degree of functionality that fit the tight form factors of space-platforms and withstand harsh environments. With the KALEXUS experiment, we have built a laser system technology demonstrator that incorporates core technologies of a space-based reference laser system for atomic physics experiments. The KALEXUS experiment is based on two narrow linewidth ($< 100$~kHz at $100~\mu$s measurement time) extended cavity diode lasers (ECDLs)~\cite{Luvsandamdin:14} that emit light at the potassium ($^{39}$K) D$_2$ transition wavelength of $766.7$~nm. The ECDLs have specifically been designed and built for applications in cold atom experiments on sounding rockets. An absolute frequency stabilization of one laser on the atomic transition of $^{39}$K by means of frequency modulation spectroscopy (FMS)~\cite{Bjorklund:80} is combined with a relative frequency stabilization using a beat note measurement between the two lasers. Hardware tests and flight demonstration of this frequency stabilized laser system are the main goal of the KALEXUS experiment. We combine these hardware tests with two additional features: autonomy and redundancy. Both are specifically important for space applications where experimenters can not easily interact with the experimental setup. We implemented an autonomy concept that automates the steps required to achieve a frequency stabilized laser system for cold atom experiments in a non-laboratory environment. For the redundancy concept, hardware and software architectures are combined to implement backup and switching options that reduce single points of failure. Thus, KALEXUS is a pathfinder experiment for ECDL technology and automated laser stabilization in space. Various applications will benefit from an increased experience and technology readiness level (TRL)~\cite{TRLESA:2016,TRLESAHandbook:2008} of such laser systems. 

The overall setup of the KALEXUS experiment and its main subsystems are described in Section~\ref{sec:experimentalsetup}. An analysis of the experiment performance during the flight is presented in Section~\ref{sec:results}. Section~\ref{sec:summary} gives a summary and outlook.  

\section{Experimental Setup}
\label{sec:experimentalsetup}
The KALEXUS experiment autonomously frequency stabilizes two ECDLs on a sounding rocket using FMS, and the relative frequency difference between the lasers. 

Sounding rockets -- as other space platforms -- have strong requirements regarding the size and the robustness of their payload (for a description of the TEXUS sounding rockets, see e.g.~\cite{TEXUS:2016,ESASoundingrockets:2016,Franke:2001}). Due to the limited space inside the rocket, the payload has to be compact and fit inside the available cylindrical volume. It also needs to withstand strong vibrations and accelerations during liftoff, reentry into the atmosphere, and landing. A typical profile of the accelerations during launch of the two-stage VSB-30 rocket motor that is used in the TEXUS sounding rocket program can be found in~\cite{Palmerio:2005}, overall flight duration and altitude are given in~\cite{Garcia:2011}. On sounding rockets, the microgravity time available for experiments is generally limited to a few minutes, which means that complete functionality of the experiments needs to be restored quickly after launch.

The KALEXUS setup is designed to fulfill these requirements. It uses a modular architecture consisting of a laser module, a spectroscopy module, and an electronics module. Each module has an individual housing with optical and electrical interfaces. A picture of the assembled flight system is shown in Fig.~\ref{fig:KALEXUS_beschriftet}. To counteract temperature increase due to excess heat from the electronics during the flight, a heat sink made from an aluminum block with copper pipes is placed underneath the main experiment modules. Up to several seconds before launch, the heat sink is water cooled to 18~$^{\circ}$C, and provides a limitation of the temperature increase of the laser module to 3~K during the whole flight. The heat sink is mounted to a base plate that fits inside the TEXUS sounding rocket, which confines the experiment diameter to 375~mm. The experiment has a height of 186~mm and a mass of 16~kg. It is enclosed by a pressurized metal dome at $1.2$~bar. The system was subject to thermal, pressure, and electrical interface tests, and the hardware was vibration tested at $8.1$~\textit{g}$_{RMS}$ to qualify for flight. 
\begin{figure}
	\centering
		\includegraphics[width = 0.95\columnwidth]{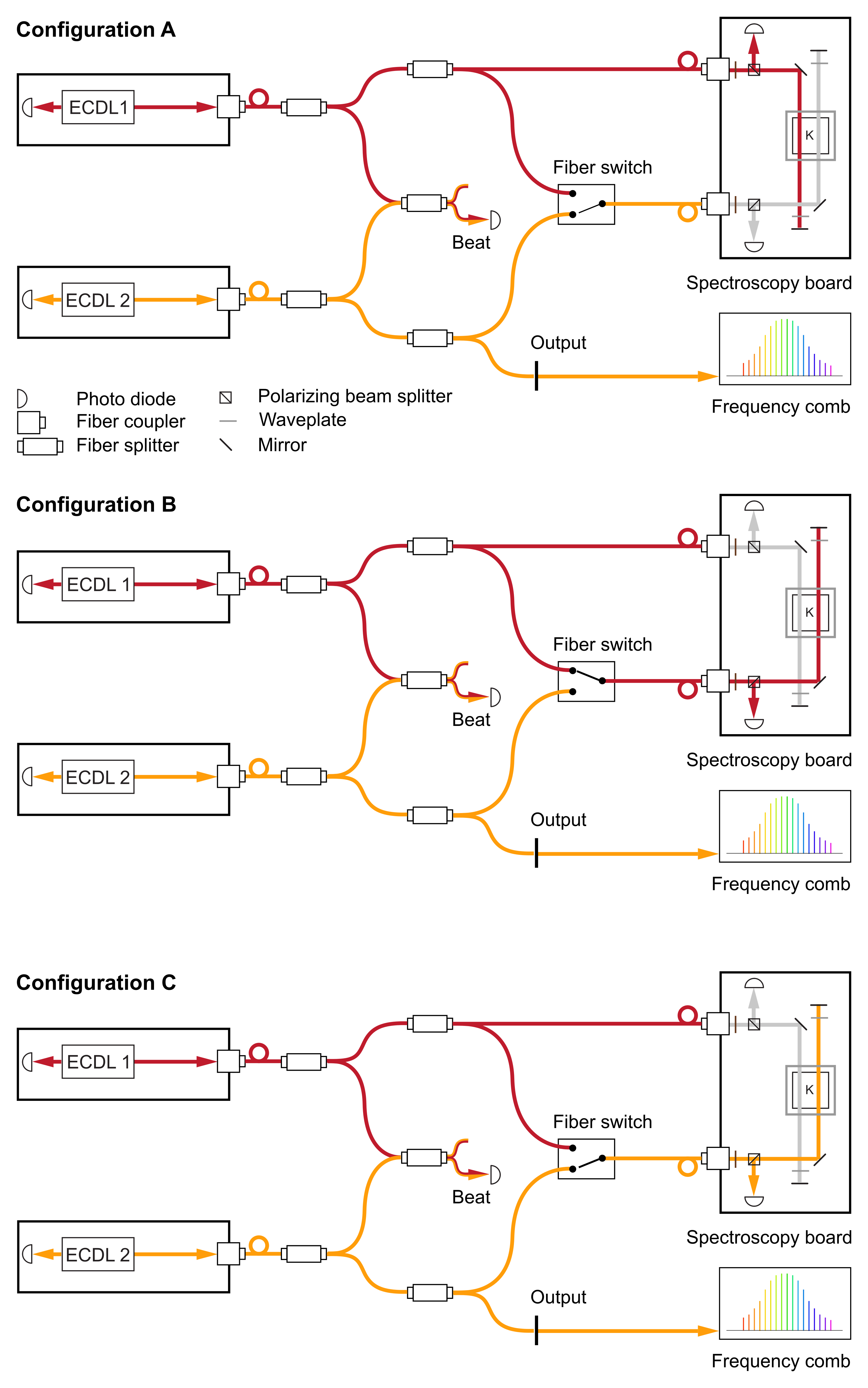}
		\caption{Schematic of the optical layout: light from two ECDLs is fiber coupled and guided via different paths to the spectroscopy module and the beat note detection. The optical paths can be switched between three different configurations (A,~B,~C, from top to bottom), so that the path through the spectroscopy module can be changed and the methods of frequency stabilization can be swapped between the lasers. This is part of the implemented redundancy concepts. Light from ECDL~2 also leads out of the KALEXUS dome for external frequency measurements via an optical interface: it is connected to the optical frequency comb by Menlo Systems that is part of the FOKUS experiment~\cite{Lezius:2016,Giunta:16}.}
	\label{fig:Optical_path}
\end{figure}

The optical layout of the experiment is shown in Figure~\ref{fig:Optical_path}. Fiber coupled light is split off from both lasers and overlapped on a fast photodiode for beat note detection. Light from each laser is also guided to the spectroscopy board via an optical switch. Additionally, light from ECDL~1 bypasses the optical switch and is directly connected to the spectroscopy board, while light from ECDL~2 is split off for external detection. 

With the spectroscopy setup, one laser can be frequency stabilized by means of FMS (spectroscopy lock). The second laser is then frequency stabilized with a fixed offset relative to the first one using the beat note frequency measurement (offset lock). By using the incorporated redundancy architecture, the frequency stabilization method assignment can be exchanged between ECDL~1 and ECDL~2. The redundancy hardware includes the optical switch, synchronized RF-switches, and a dual spectroscopy setup. In total, three different optical path configurations are possible: configurations A, B and C. An overview is shown in Figure~\ref{fig:Optical_path} and summarized in Table~\ref{tab:OpticalPathConfigurations}. 

\begin{table}
	\centering
		\begin{tabular}{l | l | l|l}
			Config.&ECDL~1& ECDL~2& Optical Switch\\
			\hline
			A						& Spectroscopy & Offset & bypassed \\
			B						& Spectroscopy & Offset & used \\
			C						& Offset & Spectroscopy & used \\
		\end{tabular}
	\caption{Optical path configurations.}
	\label{tab:OpticalPathConfigurations}
\end{table}

In the default configuration (A), light from ECDL~1 that bypasses the optical switch is used for the spectroscopy lock. For the other two configurations, light that enters the spectroscopy board through the optical switch is used for the spectroscopy lock, either from ECDL~1 (config. B) or from ECDL~2 (config. C). In each configuration, the laser that is not spectroscopy locked is stabilized with an offset lock. 

During flight, the experimental control switches between the three beam paths and stabilizes the lasers each time. Stabilization of both lasers as well as tests of this redundancy architecture are performed completely autonomously. The following sections will describe the experiment's subsystems and the experimental control. 

\subsection{Laser system}
The laser system consists of two micro-integrated ECDLs that emit light at the $^{39}$K D$_2$ transition wavelength of $766.7$~nm. ECDLs are often used in ground based precision experiments in which a narrow linewidth and low frequency noise are of advantage, e.g. for phase-locked lasers (\cite{Schmidt:2011}, see also~\cite{LeGouet2007}). In the context of space applications, mechanical stability is an additional, important factor. Here, monolithic diode lasers -- distributed feedback (DFB) or distributed Bragg reflector (DBR) -- have been favored due to their intrinsic mechanical stability. In addition, the same consortium as involved in the KALEXUS experiment has already carried out successful technology tests with DFB lasers as light sources in drop tower experiments~\cite{Schiemangk:15} and on sounding rockets~\cite{Lezius:2016,Giunta:16,Schkolnik:2016}. Since these types of lasers come with a broader linewidth and higher frequency noise compared to ECDLs~\cite{Lewoczko-Adamczyk:15}, future scientific experiments would benefit from space-qualified ECDL technology. Recently, compact and robust ECDLs have been developed to enable their use for space applications. The KALEXUS laser system is based on this type of laser, of which a thorough description and characterization can be found in~\cite{Luvsandamdin:14}. 

\begin{figure}
	\centering
		\includegraphics[width=0.95\columnwidth]{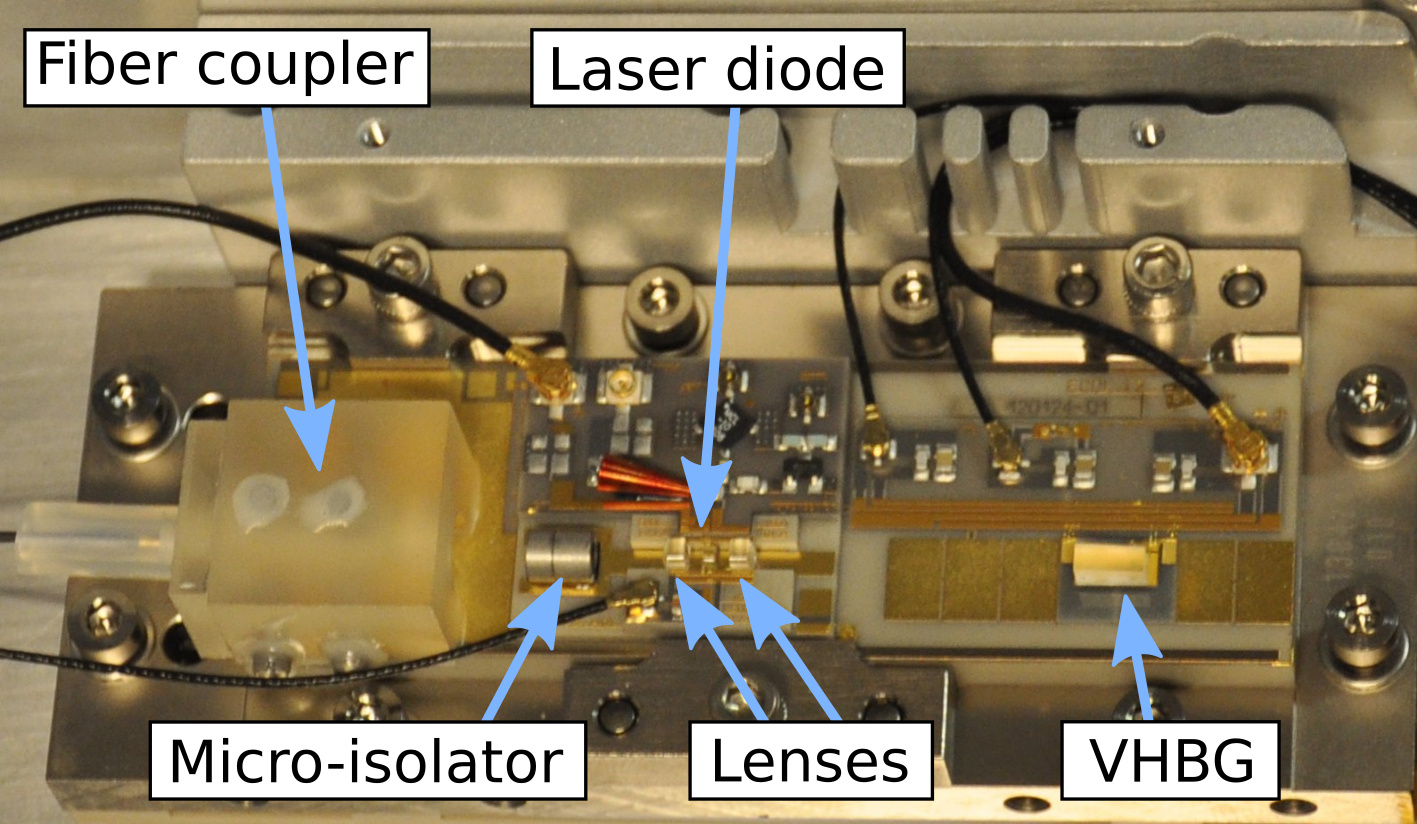}
	\caption{One of the fiber coupled ECDLs. The microbench has a footprint of $25 \times 80$~mm$^2$. Two ECDLs are inside the laser module (not shown here) with outer dimensions of $145 \times 195 \times 42$~mm$^3$. }
	\label{fig:Laser}
	\vspace{0.5cm}
	\centering
		\includegraphics[width=0.95\columnwidth]{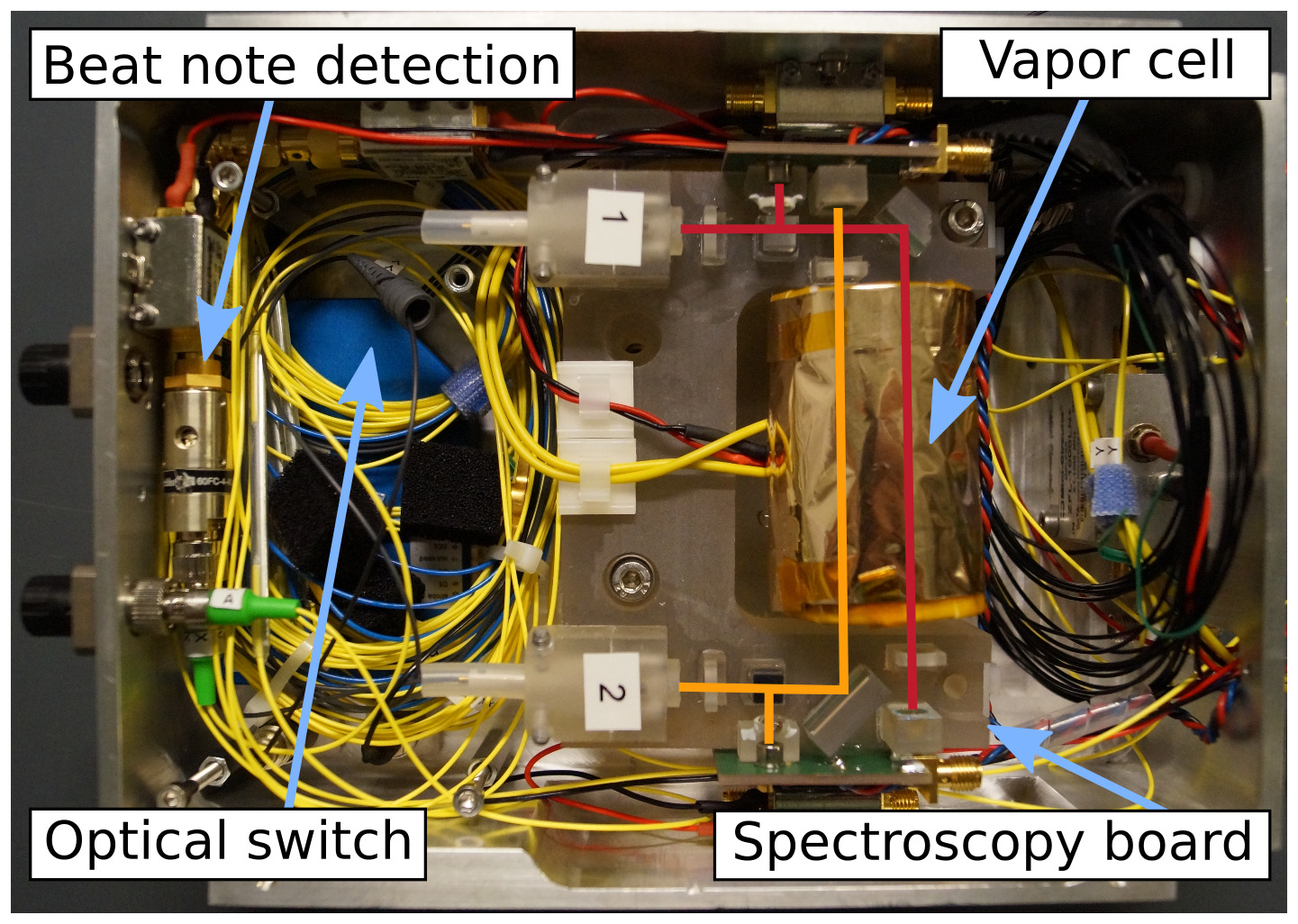}
	\caption{Spectroscopy module ( $145 \times 195 \times 67$~mm$^3$) housing the optical switch, beat note detection, and spectroscopy board. The spectroscopy board consists of a Zerodur optical bench with a heated K vapor cell and an optical setup for FMS.}
	\label{fig:Spectroscopy_Module}
	\vspace{0.5cm}
\centering
		\includegraphics[width=0.95\columnwidth]{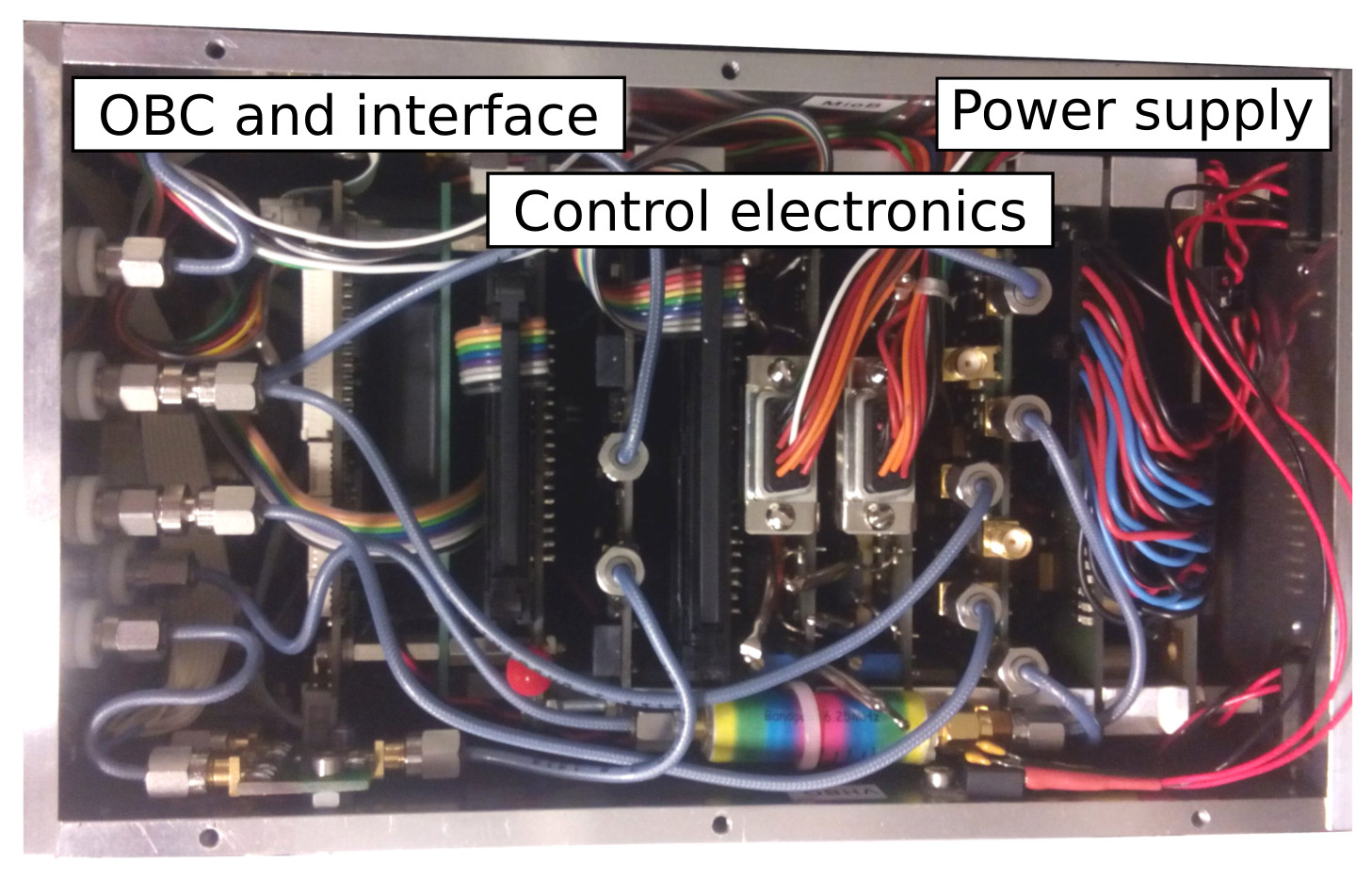}
	\caption{Electronics module with the on-board computer (left) and the control electronics (right). Included are cards to control the laser currents, laser temperatures, laser frequency and additional cards for detection of photodiode signals, power supply and interface to the OBC. The electronics module has dimensions of $128 \times 218 \times 157$~mm$^3$.}
	\label{fig:Electronics}
	
\end{figure}

Each laser has an aluminum nitride micro-optical bench as base plate, on which a ridge-waveguide laser diode, micro-optics, and micro-electronics are integrated, see Figure~\ref{fig:Laser}. Light exits from the front and the rear of the laser diode. The extended cavity is formed by the front facet of the laser chip and a volume holographic Bragg grating (VHBG), which is positioned at the rear of the laser module. 
On the front side of the laser diode and after passing a micro-isolator, the main beam is fiber coupled directly on the micro-optical bench using a Zerodur~\cite{Zerodur:2016} fiber coupler. The fiber coupling technique is described in~\cite{Duncker:14}. With coupling efficiencies around $65\%$, we have about~$15$~mW available ex fiber at maximum current. A free space photodiode on a circuit board behind the laser module monitors the optical power of the rear beam of each laser. Each laser is temperature stabilized with an individual peltier element for laser body and a micro-thermoelectric cooler for its VHBG, with temperature setpoints near $28$~$^{\circ}$C. The lasers operate at currents between 150 and 236~mA. However, the exact values depend on the software detection that is part of the automated frequency stabilization. 

Before entering the spectroscopy board for frequency stabilization (see Figure~\ref{fig:Optical_path}), the light of the two lasers passes through a fiber based distribution system inside the spectroscopy module, which is shown in Figure~\ref{fig:Spectroscopy_Module}. 

\subsection{Spectroscopy}
The spectroscopy board itself (see Figure~\ref{fig:Spectroscopy_Module}) is based on an optical bench made from Zerodur~\cite{Zerodur:2016} -- as described in~\cite{Duncker:14} -- with a footprint of $100 \times 75~$mm$^2$ and a thickness of~$35$~mm. Using the glass ceramic Zerodur as master-material has the advantage of a negligible temperature-induced expansion ($0 \pm 0.02 \times 10^{-6}$~K$^{-1}$~\cite{Zerodur:2016,Duncker:14}) over a broad range of temperatures. With monolithic optics and adhesive bonding the setup gains further stability. 

Two separate beam paths for saturated absorption spectroscopy are necessary for the redundancy architecture described above. They are realized on the spectroscopy board (Figure~\ref{fig:Spectroscopy_Module}), as shown schematically in Figure~\ref{fig:Optical_path}. The two paths use separate beam guiding optics, but share one vapor cell for volume optimization. Pump and probe beams for FMS are realized by retro-reflecting the incoming beam behind the vapor cell. To increase the vapor pressure in the cell, it is heated to a temperature of about $40^{\circ}$~C. By reading out the photodiode behind the vapor cell, the FMS signal can be used to stabilize the emission frequency of the laser to the $^{39}$K $F=1/2 \rightarrow F'$~crossover line of the D$_2$ transition. 
    
\subsection{Electronics}
The electronics module houses an on-board computer (OBC) and the control electronics for the entire experiment, which are shown in Figure~\ref{fig:Electronics}. A single power source is needed for the experiment during the flight. Voltage conversion and power distribution happens inside the electronics module. The power consumption of the whole KALEXUS system lies around $43$~W. 

For the control electronics, a modular system of 10 individual cards with specific functionality is combined to form a stack that is connected via a bus system. Each card has a footprint of $100 \times 100$~mm$^2$ and a thickness of around $17$~mm. The stack contains cards to control the temperatures and currents of the two ECDLs, count the frequency of the beat note between them, read in several photodiode signals for monitoring, and stabilize the lasers to the spectroscopy signal and the beat note frequency~\cite{LUHElektronik}. The complete software for autonomous experiment control runs on the OBC, which is a Kontron MOPSlcdLX. It is connected to the control electronics and supported by FPGAs on the control electronics for time-critical jobs. 

\subsection{Experiment control}
\label{sec:expcontrol}
Experiment control is incorporated in the overlaying structure of a time-dependent experimental sequence. Within the sequence, the software autonomously controls individual hardware components, e.g. through laser current adjustment or optical switching. The experimental sequence is divided into different parts: there are two phases (phase~1 and phase~2) with different experimental objectives during microgravity, as well as additional parts for preparation, launch, reentry and landing, in order to optimize the experiment and to protect the payload from damage. Phase~$1$ of the experimental sequence in microgravity tests autonomous frequency stabilization and redundancy architecture. Here, the optical path configuration is switched between all three possible options (A, B, C) and frequency stabilization of both lasers is realized each time. In phase~$2$, the lasers are constantly stabilized in the default configuration, and laser parameters and beat note frequency are monitored. The beat note frequency is also monitored externally through comparison with the FOKUS~\cite{Lezius:2016,Giunta:16} experiment launched on the same rocket. To fall inside the range required for the external frequency comparison, the target value for the offset frequency is kept constant at $464$~MHz throughout the flight. 

\begin{figure*}[th]
	\centering
		\includegraphics[width = 0.95\textwidth]{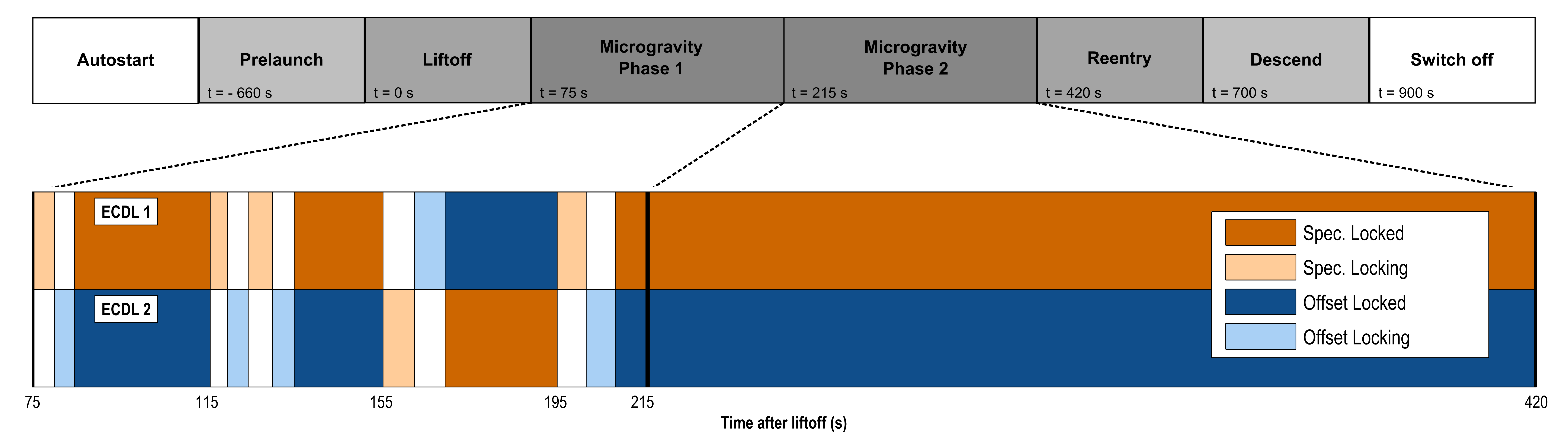}
	\caption{Top: The different states of the experiment before and throughout the flight. Bottom: ECDL~1 and ECDL~2 in microgravity during flight. Here, each laser is in one of four modes: 1. Frequency locked on spectroscopy signal (dark orange), 2. Currently trying to spectroscopy lock (light orange), 3. Frequency locked on beat note frequency (dark blue), 4. Currently trying to offset lock (light blue). During microgravity, the different optical paths result in a swap of the laser function from spectroscopy lock to offset lock and vice versa, at $t = 155$~s and $t = 195$~s. White indicates that no locking information is accessible: during locking of one laser, the system is occupied and can not provide information about the other laser.}
	\label{fig:timeline_results}
\end{figure*}

The autonomy concept was implemented in form of a multi-level state-machine. After switch-on, the OBC automatically starts the autonomous control software and enters the experimental sequence. The program uses the first level of the state-machine -- the experimental state-machine -- to determine when to go from one part in the sequence to the next. On this level, our software reacts to signals that it receives from the TEXUS system. The signals are programmed at specified times relative to liftoff and have been determined using calculated flight parameters, such as microgravity start and end. Figure~\ref{fig:timeline_results} (top) shows the experimental sequence with the relevant experimental states, illustrating the first level of the state-machine. 

Within each state of the experimental sequence, the second level of the state-machine -- the locking state-machine -- automatically stabilizes the emission frequencies of both lasers using an automated locking mechanism, and evaluates the frequency locks. The automated locking mechanism is applied to both, the spectroscopy and the frequency-offset stabilization. It follows three main steps: first a pattern detection, then a fine-tuning algorithm, and finally onset of the feedback loop. The program starts by stabilizing one laser using the spectroscopy locking technique. For the first step of the laser stabilization, the current is changed coarsely until the standard deviation of the FMS error signal -- normalized for noise -- passes a threshold value, which indicates atomic absorption and thus roughly correct frequency. In the second step, the FMS error signal is compared to a stored reference signal. Here, a cross-correlation function is used to determine the frequency offset (algorithm based on~\cite{Bartosch:2013}). Subsequently, the injection current is adjusted in order to match the two signals. Once the adjustment is complete, i.e. the laser's frequency scan is centered around the desired absorption peak, the stabilizing feedback loop is activated in the final step. For the offset-frequency stabilization, the same three steps are used, but the input signals and corresponding threshold criteria are different: e.g. for pattern detection, the laser frequency scan has to yield a linear beat note frequency signal with a specific slope. The locking state-machine monitors whether both lasers are stabilized by analyzing the error signals, beat note frequency, and feedback currents. If required, frequency stabilization with the automated locking mechanism is re-initialized automatically. 

In order to combine redundancy architecture and autonomy concepts, the selection of redundancy hardware options (i.e. switching optical paths) is incorporated into the software: while in microgravity, the locking state-machine monitors the time it takes to frequency stabilize a laser. If the laser cannot be frequency stabilized within a specified time, the program automatically switches to the next beam path configuration (see Table~\ref{tab:OpticalPathConfigurations}). This way, the time that is potentially lost due to malfunctioning hardware is limited. 

Laser monitoring and housekeeping data as well as state-machine information is stored on the OBC flash disk. Every $3$~seconds, a reduced data set is sent to ground via a serial antenna link and displayed on the ground control station for monitoring. To allow manual override from ground in the case of an emergency, a limited set of commands to control the experiment (e.g. change laser currents, switch between beam path configurations A, B and C, switch off lasers) was available. 

\section{Results}
\label{sec:results}
On 23rd of January 2016, the KALEXUS experiment was launched on-board the TEXUS~53 sounding rocket from the Esrange Swedish Space Center near Kiruna, Sweden. Both lasers could be stabilized in microgravity and the tests of redundancy concepts were successful. With the software for autonomous experiment control fully functioning, we did not use any manual experiment control throughout flight. The following sections firstly describe the experiment sequence and secondly the frequency stabilization during flight. 

\subsection{Flight}
During the TEXUS~53 sounding rocket flight, the peak acceleration along the direction of flight ($z$-axis) reached $12.1$~\textit{g} during launch 
and up to $15.7$~\textit{g} along the other axes during reentry~\cite{OHBquick}. With a maximum altitude of $252.6$~km, TEXUS~53 achieved $367$ seconds of microgravity (~$<10^{-4}$~\textit{g}). 

The timeline of the experiment sequence during flight is shown schematically in Figure~\ref{fig:timeline_results} (top). As soon as the KALEXUS experiment was switched on, it initiated laser thermalization, frequency stabilization, monitoring and prepared the system to be ready for liftoff. Before launch, at $t=-660$~s, the prelaunch state started and data recording was activated. With liftoff at $t=0$~s, vibrations set in during motor ignition and ascend flight. To protect our hardware, the state-machine was programmed to limit the experiment's functionality during launch by disabling the optical switch. Microgravity was reached at $t=73$~s and lasted until $t=440$~s. Microgravity phases~1 and~2 were the main testing time, in which different experimental sequences were activated and redundancy and frequency stabilization tests were performed. Strong vibrations also occurred during reentry in the atmosphere from around $t=460$~s, which ceased after parachute deployment at $t=620$~s. From that point onwards, the payload slowly descended to the ground, where it landed at $t=888$~s. The OBC recorded data from several minutes before liftoff until right after landing ($t=900$~s), when the experiment received the signal to switch off. 

\begin{figure}[ht]
	\centering
		\includegraphics[width = 0.95\columnwidth]{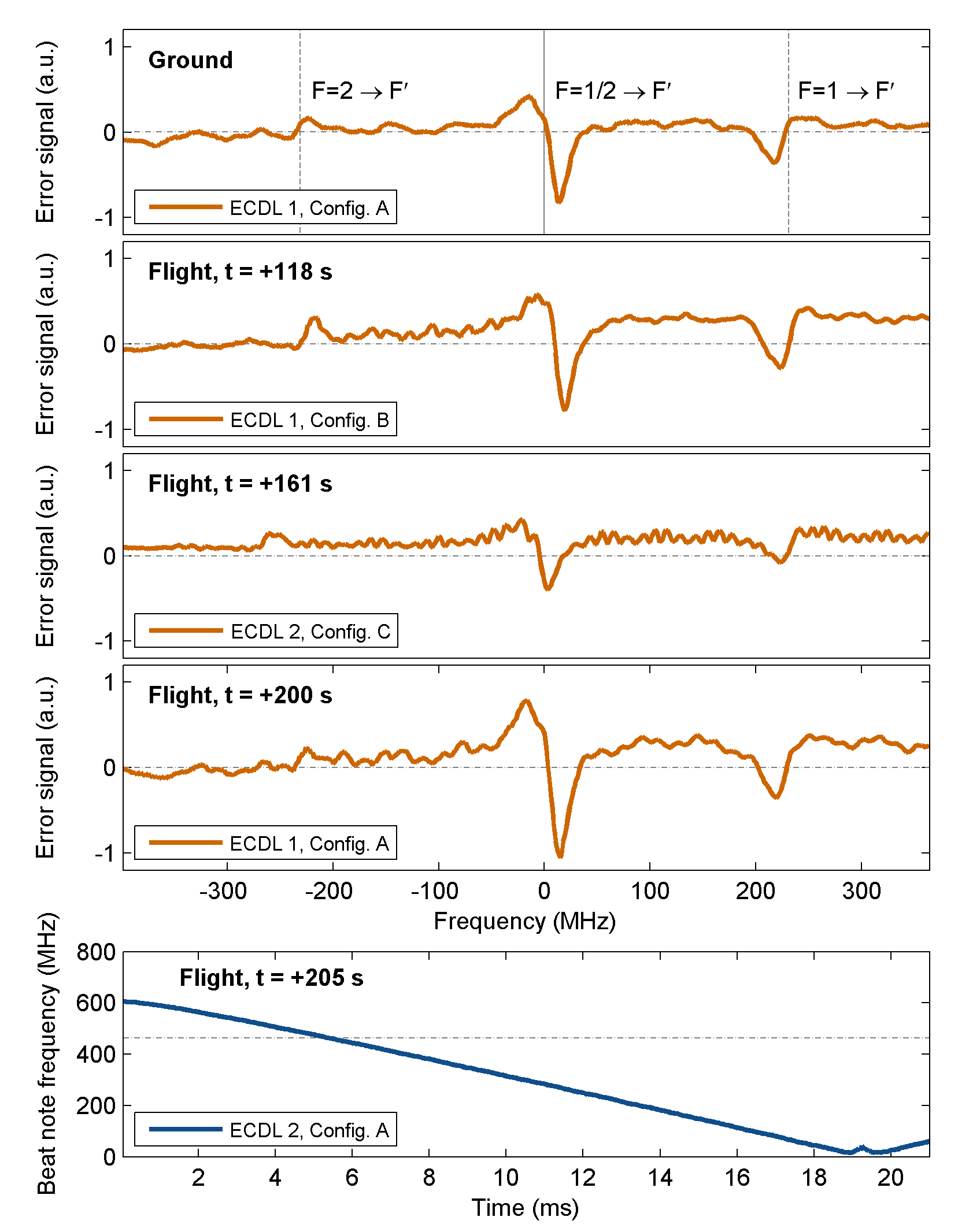}
	\caption{Absorption spectroscopy error signal of $^{39}$K on ground for optical path configuration~A and during flight for all three configurations,~A,~B~and~C. The horizontal dashed lines indicate the lock points of the signal, i.e. the setpoint for the error signal feedback loop for the spectroscopy lock. D$_2$~transitions and crossover of $^{39}$K are labeled. The bottom graph also shows a scan of the beat note signal that is used for offset locking, with the dashed line indicating the desired offset frequency (here: 464~MHz).}
	\label{fig:Kalexus_Flight_Spectroscopy_Signal}
\end{figure}

\subsection{Frequency stabilization during flight}
Both lasers were switched on throughout the whole flight until the experiment was powered off, and we had automated frequency stabilization activated at all times. Due to the vibrations during launch, the frequency stabilization was disturbed and had to be re-initiated by the locking state-machine. Several seconds after the vibrations subsided, both lasers could be stabilized. Here, we focus on describing the frequency stabilization in microgravity, after launch, since this will be the relevant experimental environment in future science missions. Results of the laser stabilization monitoring for the duration of microgravity are shown in the bottom part of Figure~\ref{fig:timeline_results}. ECDL~1 is shown in the top and ECDL~2 in the bottom part of the graph. 

At $t=75$~s, phase~1 of the KALEXUS experiment began, in which the automated frequency stabilization and redundancy hardware were tested. We implemented a mandatory re-initialization of the locking mechanism when microgravity set in, which is reflected in the data. At times $t_B=115$~s, $t_C=155$~s and $t_A=195$~s, the state-machine received programmed signals to actively change the optical path and re-stabilize the lasers by initializing the stabilization algorithm. The spectroscopy stabilization was activated first, the offset stabilization followed afterward. At all three times, we observed successful stabilization of both lasers. In two cases, it took one attempt and in one case ($t_B=115$~s) it took two attempts for successful stabilization. We recorded the respective saturated absorption error signals of $^{39}$K on ground and during flight at three different times during microgravity (in all three optical path configurations), which are shown in Figure~\ref{fig:Kalexus_Flight_Spectroscopy_Signal}. The scan frequency is around $23$~Hz. Also shown is a scan of the beat note signal that was used for offset locking. 
The spectra observed on ground and during flight agree qualitatively. Deviations in signal amplitude are due to differences in the optical power among the different paths. 
The small oscillations visible in the error signals recorded during flight have also been observed on ground and are most likely due to an etalon effect in the system, which is currently under investigation as part of the post-flight analysis.  

Phase~2 of the KALEXUS experiment started at $t=215$~s. During phase~2, both lasers were kept stabilized and parameters such as laser temperatures, laser currents, beat note frequency and error signal were monitored to test longer timescale behavior of the lasers and potential drifts. We did observe a constant increase in feedback current for the laser stabilization over time, which relates to an overall temperature increase of the hardware during flight. However, over the complete remaining microgravity duration of $225$~s, we observed an uninterrupted spectroscopy lock of ECDL~1 and offset lock of ECDL~2. For additional monitoring, a beat note measurement was performed throughout the flight between ECDL~2 and an external frequency comb by Menlo Systems in a separate payload on the same rocket~\cite{Lezius:2016,Giunta:16}. This beat note measurement confirmed independently that both lasers were frequency stabilized during phase~2. 

During reentry -- as during launch -- the high vibrations disturbed the frequency stabilization, and the locking mechanism was engaged. After acceleration decreased, both lasers could be re-stabilized. The hardware survived the flight and was fully functioning after recovery. 

\section{Summary and Outlook}
\label{sec:summary}
With the KALEXUS experiment, we have built a laser system technology demonstrator on a sounding rocket. Its core technologies are micro-integrated ECDLs, a spectroscopy board made of Zerodur, compact electronics, and an autonomously operating control software. The laser system was designed and built with future space-borne setups for cold atom quantum sensors in mind. To demonstrate functionality in a space environment, KALEXUS was launched on-board the TEXUS~53 sounding rocket. During flight, all experimental sequences were performed autonomously. These included automatic frequency stabilization of two micro-integrated ECDLs at $766.7$~nm with two different methods: one laser was frequency stabilized with an absorption spectroscopy signal of $^{39}$K while the second laser was stabilized using the beat note frequency of the two lasers. Both lasers stayed frequency stabilized over the designated testing duration of $225$~s in microgravity. In addition, hardware and software redundancy concepts were successfully tested during flight. The flight proven system thus achieved a TRL of 9 for sounding rocket application, following the ESA classifications~\cite{TRLESA:2016,TRLESAHandbook:2008}. 

With this level of technological maturity, we can aim to implement the KALEXUS technologies into more complex experimental setups, e.g. to push for a space-based EEP test. In this context, ECDL technology is now a flight-proven option for applications requiring narrow linewidth light sources, such as cold atom experiments similar to those described in~\cite{Herrmann:2012,Muntinga:2013,Schlippert:2014}. The single and dual species interferometry experiments on sounding rockets that are currently in preparation~\cite{Zoest:2010b,Stamminger:2015,Schkolnik:2016} are therefore no longer restricted to using rocket qualified DFB lasers~\cite{Lezius:2016,Giunta:16,Schkolnik:2016}. Science missions on satellites, such as the proposed STE-QUEST~\cite{Aguilera:2014}, require a qualification of the KALEXUS technology beyond sounding-rockets~\cite{ESASTEQUEST:2012,TRLESAHandbook:2008}. Additional environmental tests necessary for satellite-based science missions include exposure to radiation and thermal variations. These in combination with ongoing parallel technological developments for small satellites~\cite{Oi:2016} and ISS-based experiments with cold atoms~\cite{Leveque:2014,Leveque:2015,CAL:2016} will increase the available hardware for atomic physics and quantum optics experiments in space. As part of these developments, the sounding rocket qualification of the laser system technology achieved with the KALEXUS experiment was an important step towards application in space-based quantum sensors and beyond.  

This work is supported by the German Space Agency DLR with funds provided by the Federal Ministry for Economic Affairs and Energy under grant number DLR 50WM1345. 

We would like to thank C.~Grzeschik for helpful discussions, M.~Mihm for support with the integration of the optical system and Prof.~Klaus Sengstock for providing the facilities for Zerodur technology integration at the Universit\"at Hamburg. We would also like to thank Airbus~DS for technical support throughout the mission, as well as SSC, MORABA and OHB for assistance during the launch campaign.  



\bibliographystyle{ieeetr}

\bibliography{KALEXUS-Arxiv}

\begin{thebibliography}{10}
\bibitem{LUHElektronik}
All cards for experiment control have been developed by Leibniz-Universit\"at
  Hannover for previous laser control and cold atom experiments, but were
  adjusted to match the required specifications of the {KALEXUS} experiment.
\end{thebibliography}

\end{document}